\documentclass[12pt]{article}
\usepackage{epsfig}
\usepackage{psfrag}
\usepackage{latexsym}
\usepackage{amssymb}
\usepackage{amsmath}
\usepackage{mathrsfs}

\def\be{\begin{equation}}
\def\ee{\end{equation}}
\def\bc{\begin{center}}
\def\ec{\end{center}}
\def\bea{\begin{eqnarray}}
\def\eea{\end{eqnarray}}
\newcommand{\ba}{\begin{array}{c}}
\newcommand{\bad}{\begin{array}{ccc}}
\newcommand{\ea}{\end{array}}
\def\nn{\nonumber}
\def\dd{\displaystyle}

\begin{document}
\begin{titlepage}
\vspace*{-1cm}
\phantom{hep-ph/***} 

\hfill{DFPD-08/TH/04}

\vskip 2.5cm
\begin{center}
{\Large\bf A predictive $A_4$ model, Charged Lepton Hierarchy\\
\vskip .2cm
and Tri-bimaximal Sum Rule}
\end{center}
\vskip 0.2  cm
\vskip 0.5  cm
\begin{center}

{\large Yin Lin}~\footnote{e-mail address: yinlin@pd.infn.it}
\\
\vskip .1cm
Dipartimento di Fisica `G.~Galilei', Universit\`a di Padova 
\\ 
INFN, Sezione di Padova, Via Marzolo~8, I-35131 Padua, Italy
\\
\end{center}
\vskip 0.7cm
\begin{abstract}
\noindent
We propose a novel $A_4$ model in which the Tri-Bimaximal (TB) neutrino mixing
and the charged lepton mass hierarchy are reproduced simultaneously.
At leading order, the residual symmetry of the neutrino sector is
$Z_2 \times Z_2$ which guarantees the TB mixing without 
adjusting ad hoc free parameters.
In the charged lepton sector, 
one of the previous $Z_2$ is maximally broken
and the resulting mass matrix is nearly diagonal and hierarchical.
A natural mechanism for the required vacuum alignment is given
with the help of the supersymmetry and an abelian symmetry factor.
In our model, subleading effects which could lead to
appreciable deviations from TB mixing are very restrictive giving rise to
possible next-to-leading predictions.
From an explicit example, we show that  our ``constrained'' $A_4$ model
is a natural framework,
based on symmetry principle, to incorporate the TB sum rule:
$\text{sin}^2\theta_{12}=1/3+2\sqrt{2}~(\text{cos} \delta ~\text{sin}\theta_{13})/3~.$
\end{abstract}
\end{titlepage}
\setcounter{footnote}{0}
\vskip2truecm

\section{Introduction}

Nowadays continuous improvement on the knowledge of neutrino oscillation parameters  
makes desirable a neutrino model building going beyond the mere fitting procedure.
In particular the leptonic mixing pattern, so different from the one in the quark sector, provides a 
non-trivial theoretical challenge. The present data \cite{data}, at 1$\sigma$:
\be
\theta_{12}=(34.5\pm 1.4)^o~~~,~~~~~~~\theta_{23}=(42.3^{+5.1}_{-3.3})^o~~~,~~~~~~~\theta_{13}=(0.0^{+7.9}_{-0.0})^o~~~,
\label{angles}
\ee
are fully compatible with the so-called Tri-Bimaximal (TB) mixing matrix:
\be
U_{\text{TB}}=\left(
\begin{array}{ccc}
\sqrt{2/3}& 1/\sqrt{3}& 0\\
-1/\sqrt{6}& 1/\sqrt{3}& -1/\sqrt{2}\\
-1/\sqrt{6}& 1/\sqrt{3}& +1/\sqrt{2}
\end{array}
\right)~,
\label{TB}
\ee
which corresponds to
\be
\sin^2\theta_{12}=\frac{1}{3}~~~(\theta_{12}=35.3^o)~~~,~~~~~~~\sin^2\theta_{23}=\frac{1}{2}~~~,~~~~~~~\sin^2\theta_{13}=0~~~.
\ee
Several interesting ideas leading to a nearly TB mixing have been suggested 
in the last years \cite{altarelli}.
The TB mixing has the advantage of correctly describing the solar mixing angle, which, at present, is the most precisely
known. Indeed, its 1$\sigma$ error, $1.4$ degrees corresponds to less than $\lambda_c^2$ radians, where $\lambda_c\approx 0.22$ 
denotes the Cabibbo angle. 

TB pattern belongs to the class of mixing textures which are independent
on the mass eigenstates.
Mass-independent mixing textures
usually exhibit an underlying discrete symmetry nature \cite{Lam}. 
It has been realized that the TB mixing matrix
of Eq.~(\ref{TB}) can naturally arise as the result of a particular vacuum alignment of scalars that 
break spontaneously certain discrete flavour symmetries.
A class of very promising models are based on $A_4$ flavour symmetry
\cite{TB1, TB2} and subsequently extended 
to the group $T' $ \cite{TB3} to cover a reasonable description also for quarks.
Despite of the success, the original  $A_4$ models proposed by Altarelli and Feruglio (AF) in \cite{TB2} 
require improvement for various reasons.
First of all, the leading order results of AF are affected by 
a large number of subleading corrections.
Even though these corrections are hopefully under control, they are totally independent and
 the model looses the possibility to go beyond the leading prediction. Furthermore the mass eigenvalues are completely unspecified by $A_4$ 
and the charged lepton mass hierarchy can be explained only by an extra 
Froggatt-Nielsen (FN) \cite{FN} $U(1)_{FN}$  factor.
Finally, $A_4$ alone seems unfavorable to accommodate 
quark masses and eventually be embedded into a GUT theory, despite some recent attempts in
this direction: Pati-Salam \cite{GUT1}, SU(5) \cite{GUT2, GUTAF}, SO(10) \cite{GUT3}.
Indeed, it is a quite non trivial task to reproduce all
charged fermion hierarchies compatible with a natural mechanism for the vacuum alignment.
Particular efforts have been made in this direction.
For example, in \cite{A4-LR}, the problem of fermion hierarchy is partially solved by embedding 
$A_4$ into a continuous left-right symmetry. 
However the question of a natural vacuum alignment in their model remains open.
In the recent proposal of $A_4$ model in a 5D SUSY SU(5) GUT \cite{GUTAF}, the fermion mass
hierarchies and mixing are a result of the interplay of three different sources: 
a discrete flavour group based on $A_4$, wave-function suppressions
of bulk fields and an additional $U(1)_\text{FN}$.

In the literature, there are many mechanisms based on different discrete groups that can successfully 
describe TB mixing at leading order. Another important issue for the model building 
is if there were some criteria to distinguish various constructions.
One possibility is to go beyond the leading order prediction.
There is a remarkable sum rule in the lepton mixing sector \cite{dev}:
\be
\text{sin}^2\theta_{12}=\text{sin}^2\theta^\nu_{12}+\text{sin}^22\theta^\nu_{12} ~\text{cos} \delta ~\text{sin}\theta_{13}~,
\label{sumrule}
\ee
where $\delta$ is the Dirac CP violation phase.
This sum rule can be derived when
maximal $\theta_{23}$ and $\theta_\text{13} = 0$ 
come from the neutrino sector at leading order 
and a non vanishing $\theta_\text{13}$ arises 
only when one includes
small charged lepton mixing (under certain assumptions such as 
$\theta^e_{12}$ or $\theta^e_{13}$ dominance as will be explained in the section 5.1).
In flavour models in which $\text{sin}^2\theta^\nu_{12}=1/3$ holds almost exactly, 
this sum rule should be a precise test.
However, generally, in the dynamical realization of TB mixing pattern based 
on discrete flavour symmetries,
the Eq.~(\ref{sumrule}) is not precisely predicted. For example, in the context of the AF models,
higher order corrections generically contribute to neutrino masses as well as to
the charged lepton masses, all of relative order $\lambda_c^2$, being $\lambda_c$ the Cabibbo angle. 
Then one could only expect that
$\text{sin}^2\theta^\nu_{12}=1/3 +O(\lambda_c^2)$ and $\text{sin}^2\theta_{13}=O(\lambda_c^2)$,
 if the model is, in some sense, ``unconstrained''.

In this work, we will give a realization of the TB pattern solving the charged lepton hierarchy
problem and discuss a possible subleading prediction according to Eq.~(\ref{sumrule}).  
The model is supersymmetric and based only on a discrete symmetry $A_4 \times G $
where $G$ is an abelian factor. The mixing matrix of Eq.~(\ref{TB}) is obtained in the neutrino sector 
by the spontaneous $A_4$ breaking. 
Supersymmetry (SUSY) is introduced to simplify the discussion of the vacuum
alignment.
At the lowest order of the expansion parameters $\langle \varphi \rangle /\Lambda \ll 1$, 
only the tau mass is generated together with the TB mixing. 
The muon and electron masses are subsequently generated  by higher orders
of the same expansion, similar to what happens in the recently proposed $S_3$ model \cite{S3}. 
This is one of the distinguished features of our model:
the charged lepton hierarchy is also controlled by
the spontaneous breaking of $A_4$ without introducing an extra $U(1)_{FN}$ factor.
The abelian factor $G$ is given by $Z_3 \times Z'_3$ {\footnote {It is worth to 
keep in mind that the abelian factor $Z_3$ or $Z'_3$
does not correspond to a discrete component of $U(1)_{FN}$.
As we will see they are not directly responsible for the charged lepton hierarchy.}}.
The presence of an abelian factor $G$ is essential for our construction.
First of all, $G$ guarantees the misalignment in flavour
space between the neutrino and the charged lepton mass eigenstates, responsible
for both TB mixing and charged lepton hierarchy.
Furthermore, $G$ plays an important role in suppressing subleading contributions
in order to keep the model predictive.
From this point of view, the present $A_4$ model is more constrained than the other
$A_4$ models and we will refer it as
a constrained $A_4$ model. TB mixing in the neutrino sector holds almost exactly
and subleading corrections generically lead to interesting correlations between parameters.
In particular, corrections to the charged lepton sector
are subjected by the neutrino sum rule of Eq.~(\ref{sumrule})
and offer a possible test of our model. 

In section 2 we outline the main features of our model focusing on the symmetry breaking pattern.
We then move to solve explicitly the vacuum alignment problem in a SUSY context 
in section 3 finding a new
type of minimum of the scalar potential. 
In section 4 we construct a simple model
of leptons with the symmetry breaking pattern according to the vacuum alignment. 
In section 5, we will discuss possible predictive deviations from the TB pattern,
in particular those related to the sum rule of Eq.~(\ref{sumrule}).
Finally, in section 6, we comment on other aspects of our model and then conclude.




\section{The ingredients for an alternative realization of $A_4$}

The group $A_4$ has 12 elements and four non-equivalent irreducible representations: 
one triplet and three independent singlets $1$, $1'$ and $1''$. 
Elements of $A_4$ are generated by the two generators
$S$ and $T$ obeying the relations:
\be
S^2=(ST)^3=T^3=1~~~.
\label{$A_4$}
\ee
We will consider the following unitary representations of $T$ and $S$:
\be
\begin{array}{lll}
\text{for}~1&S=1&T=1\\
\text{for}~1'&S=1&T=e^{\dd i 4 \pi/3}\equiv\omega^2\\
\text{for}~1''&S=1&T=e^{\dd i 2\pi/3}\equiv\omega
\label{singlets}
\end{array}
\ee
and for the triplet representation
\be
T=\left(
\begin{array}{ccc}
1&0&0\\
0&\omega^2&0\\
0&0&\omega
\end{array}
\right),~~~~~~~~~~~~~~~~
S=\frac{1}{3}
\left(
\begin{array}{ccc}
-1&2&2\cr
2&-1&2\cr
2&2&-1
\end{array}
\right)~~~.
\label{ST}
\ee
The tensor product of two triplets is given by $3 \times 3 = 1+1'+1''+3_S+3_A$.
From (\ref{singlets}) and (\ref{ST}), one can easily construct all multiplication rules of $A_4$. 
In particular, for two triplets $\psi=(\psi_1,\psi_2, \psi_3)$ 
and $\varphi=(\varphi_1,\varphi_2, \varphi_3)$ one has:
\bea \label{tensorproda4}
&\psi_1\varphi_1+\psi_2\varphi_3+\psi_3\varphi_2 \sim 1 ~,\nn \\
&\psi_3\varphi_3+\psi_1\varphi_2+\psi_2\varphi_1 \sim 1' ~,\nn \\
&\psi_2\varphi_2+\psi_3\varphi_1+\psi_1\varphi_3 \sim 1'' ~,\nn
\eea
  \be
   \left( 
 \ba
 2\psi_1\varphi_1-\psi_2\varphi_3-\psi_3\varphi_2 \\
 2\psi_3\varphi_3-\psi_1\varphi_2-\psi_2\varphi_1 \\
 2\psi_2\varphi_2-\psi_1\varphi_3-\psi_3\varphi_1 \\
  \ea
  \right) \sim 3_S~, \qquad
  \left( 
 \ba
 \psi_2\varphi_3-\psi_3\varphi_2 \\
 \psi_1\varphi_2-\psi_2\varphi_1 \\
 \psi_3\varphi_1-\psi_1\varphi_3 \\
  \ea
  \right) \sim 3_A~.
  \label{tensorp}
   \ee
The group $A_4$ has two obvious subgroups: $G_S$, which is a reflection subgroup
generated by $S$ and $G_T$, which is the group generated by $T$, isomorphic to $Z_3$.

First of all, we will quickly revisit the original proposal of AF based on $A_4 \times Z_3$.
Lepton SU(2) doublets $l_i$ $(i=e,\mu,\tau)$ are assigned to the triplet $A_4$ representation, while the lepton singlets $e^c$, $\mu^c$ and $\tau^c$ are assigned to $1$, $1''$ and $1'$, respectively. 
The symmetry breaking sector consists of the scalar fields neutral under the SM gauge group: $(\varphi_T,\varphi_S,\xi)$,
transforming as $(3,3,1)$ of $A_4$.
The additional $Z_3$ discrete symmetry is needed in order to reproduce the desired alignment
and for the separation of charged lepton and neutrino sectors. 
The key feature of their model is that the minimization of the scalar potential at the leading order
leads to the following VEVs:
\begin{equation}
\langle \varphi_T\rangle\propto (1,0,0)~,~~~~~\langle \varphi_S\rangle\propto (1,1,1)~,~~~~~\langle \xi\rangle\ne 0~~~.
\label{va1}
\end{equation}

In the charged lepton sector the flavour symmetry $A_4$ is broken by $\langle \varphi_T \rangle$ down to $G_T \simeq Z_3 $ generated by $T$. At leading order in $1/\Lambda$, 
charged lepton masses are diagonal simply because
there is a low-energy $G_T$ symmetry. 
In the neutrino sector $A_4$ is broken down to $G_S \simeq Z_2$ generated by $S$.
The absence of the scalar singlets $1'$ and $1''$ in the neutrino sector implies that
the resultant neutrino mass matrix is the most general one allowed by $G_S \times G_{2-3}$
where $G_{2-3}$ is generated by
$$S_{2-3}= \left(
\begin{array}{ccc}
1&0&0\\
0&0&1\\
0&1&0
\end{array}
\right).$$
$G_{2-3} \simeq Z_2$ is the permutation symmetry of the second and third generation of neutrinos.
It is by now known that the TB structure of lepton mixing is a result of the intrinsic $G_S \times G_{2-3}$
symmetry in the basis where the charged leptons are diagonal \cite{Lam}.

In our alternative realization of $A_4$ model, the neutrino sector is unchanged at leading order.
The only difference is that the TB mixing pattern results almost exact 
(its precise meaning will be clear later) even including subleading corrections.
For the charged leptons, we would like to investigate the consequence of 
an alternative alignment of $\varphi_T$:
\be
\langle \varphi_T\rangle\propto (0,1,0) 
\label{vevphiT}
\ee
which  entirely  breaks $A_4$. Such a VEV breaks also the permutation symmetry
in a maximal way, since $$ \langle \varphi_T\rangle^t S_{2-3} \langle \varphi_T\rangle=0~. $$

From the group theory point of view, there is an essential difference between the alignment 
(\ref{va1}) and that considered in the present paper (\ref{vevphiT}). 
The first one is a subgroup-preserving direction, while the
second one breaks $A_4$ down to nothing.
Observe that a natural realization of (\ref{va1}) in the AF model requires
that the scalar potential for $\varphi_T$ and ($\varphi_S, \xi$) is actually separated.
This can be done, for example, 
by imposing different abelian charges for $\varphi_T$ and ($\varphi_S, \xi$)
under $Z_3$. 
Now consider a more general scalar potential with only one triplet $\varphi_T$ plus
eventually other singlets of $A_4$: $V(\varphi_T, \cdots )$.
Interestingly, we find that leading order SUSY vacua of the type
(\ref{vevphiT}) can never been obtained without fine-tuning if 
$V(\varphi_T, \cdots )$ is symmetric only with respect to $A_4$  
(without considering additional abelian factor $G$).
An alignment of the type (\ref{vevphiT}) that entirely breaks $A_4$
must be highly fine tuned. 
For this reason, it is of general interest to find conditions under which (\ref{vevphiT}) 
could be obtained natually.
The strategy is to enlarge the $A_4$ group by an additional abelian factor $G$.
The desired alignment can then be regarded as a subgroup-preserving direction of 
$A_4 \times G$.  

Consider an enlarged flavour group $A_4 \times Z'_3$ for charged leptons.
Now the alignment (\ref{vevphiT}) preserves the subgroup $G_{T^+} \simeq Z_3$
generated by $T^+$ defined as a simultaneous transformation of $T \in A_4$ and $\omega \in Z'_3$~,
or briefly 
\be
T^+= \left(
\bad
\omega&0&0\\
0&1&0\\
0&0&\omega^2
\end{array}
\right)~.
\label{Tprime}
\ee
In addition to $G_{T^+}$, $A_4 \times Z'_3$ has another subgroup isomorphic to $Z_3$,
not contained in $A_4$:
$G_{T^-}$ generated by $T^-$ defined as a simultaneous transformation of $T$ 
and $\omega^2$~, or briefly
\be
T^-= \left(
\bad
\omega^2&0&0\\
0&\omega&0\\
0&0&1
\end{array}
\right)~.
\label{T2prime}
\ee
Hereafter, we will use $G_T$ for the subgroup $G_T \times 1$ of $A_4 \times Z'_3$.
As we will see in a moment, from the point of view of $A_4 \times Z'_3$, 
the alignment considered in our paper has rather different implications
for charged lepton masses in comparison to the original AF model based on $A_4$.

The difference between $\langle \varphi_T \rangle\propto (1,0,0) $ and
$\langle \varphi_T \rangle \propto (0,1,0)$ can be seen in another way. 
From the tensor product given in (\ref{tensorp}), we see that, in the first case, 
$\langle \varphi_T \rangle^m $ for any integer $m \geq 1$ 
is aligned in the same direction of $\langle \varphi_T \rangle$.
Instead, in the second case, this does not happen. In fact, $\langle \varphi_T \rangle \propto (0,1,0)$ implies that $\langle \varphi_T\rangle^{3n+1} \propto (0,1,0)$,
$\langle \varphi_T\rangle^{3n+2} \propto (0,0,1)$, $\langle \varphi_T \rangle^{3n+3} \propto 
(1,0,0)$. These three directions preserve respectively  $G_{T^+}$, $G_{T^-}$ and $G_T$,
subgroups of  $A_4 \times Z'_3$ defined before. 
Considering the case $n=0$,
the three orthogonal directions generated by $\langle \varphi_T\rangle$, 
$\langle \varphi_T\rangle^2$ and $\langle \varphi_T\rangle^3$ are, in our model,
responsible for the charged lepton hierarchy.
Assigning the lepton doublets $l_i$ to the triplet $A_4$ representation and
the lepton singlets $e^c, \mu^c, \tau^c \sim 1$, we have 
the most general mass matrices invariant under $G_{T^+}$, $G_{T^-}$ and $G_T$ respectively
(we will use a convention in which the conjugate fields lie on the left hand side of
superpotentials):
\be
m^{(1)}_l= \left(
\bad
0&0&\times \\
0&0&\times \\
0&0&\times
\end{array}
\right), \qquad
m^{(2)}_l= \left(
\bad
0&\times &0\\
0&\times &0\\
0&\times &0
\end{array}
\right), \qquad
m^{(3)}_l= \left(
\bad
\times &0&0\\
\times &0&0\\
\times &0&0
\end{array}
\right).
\label{ml0}
\ee
Since the masses $m^{(1)}_l$, $m^{(2)}_l$ and $m^{(3)}_l$ arise
at order $\langle \varphi_T\rangle /\Lambda$, $(\langle \varphi_T\rangle /\Lambda)^2$
and $(\langle \varphi_T\rangle /\Lambda)^3$ respectively, 
we automatically obtain the correct hierarchy between the charged lepton masses,
$m_e \ll m_{\mu} \ll m_{\tau}$, if we assume $\langle \varphi_T\rangle / \Lambda \sim \lambda_c^2 
$, being $\lambda_c$ the Cabibbo angle. 

In the next section, we will explicitly solve the vacuum alignment problem
in the enlarged flavour group, but 
now we will give an insight into some important features of our model.
The full flavour symmetry is based on $A_4 \times Z'_3 \times Z_3$
where $A_4$ is responsible for the TB lepton mixing and the abelian factor 
$Z'_3 \times Z_3$ is important in the vacuum alignment analysis.
Furthermore, $Z'_3 \times Z_3$ suppresses almost any interactions between the neutrino 
and the charged lepton sectors.
The fields of the model, together with their
transformation properties under the flavour group, are listed in Table~\ref{transform}.
In the neutrino sector, $A_4$ is broken down to $G_S$ exactly as in the AF model. 
For charged leptons, both $A_4$ and $Z'_3$ are broken, however, their ``diagonal'' 
combination $G_{T^+}$ survives at leading order. The breaking of  
$A_4 \times Z'_3$ by $\langle \varphi_T\rangle /\Lambda$, $(\langle \varphi_T\rangle /\Lambda)^2$
and $(\langle \varphi_T\rangle /\Lambda)^3$ generates, as discussed before,
a hierarchical structure of the charged lepton mass matrix $m_l$.
However, if $m_l$ were $m^{(1)}_l+m^{(2)}_l+m^{(3)}_l$, the left handed mixings
would be small enough ($\sim O(\lambda^2_c)$), whereas large right-handed mixings necessarily
would arise.
This structure of $m_l$ is disfavored by a future GUT embedding based on SU(5), since
the relation $m_l \sim m^T_d$ would imply large mixings in the down quark sector.
In general one should not expect that the large down quark mixings could be cancelled
by large up quark mixings in the context of SU(5). 
This potential problem should not be present if we assign different $Z'_3$ 
charges for lepton singlets in such a way that
$\tau^c$ couples with $\varphi_T$, $\mu^c$ with $\varphi_T^2$ and
$e^c$ with $\varphi^3_T$ as shown in Table~\ref{transform}.
In fact, we will arrive at a diagonal and hierarchical $m_l$:
\be
m_l=\left(
\bad
\varphi^3_T/\Lambda^3&0&0\\
0&\varphi^2_T/\Lambda^2&0\\
0&0&\varphi_T/\Lambda
\end{array}
\right) v_d~.
\label{mlgen}
\ee

\begin{table}
\centering
\begin{tabular}{|c||c|c|c|c||c|c|c|c|c|c|c||c|c|c|c|}
\hline
{\tt Field}& l & $e^c$ & $\mu^c$ & $\tau^c$ & $h_u$ & $h_d$& 
$\varphi_T$ &  $\xi'$ & $\varphi_S$ & $\xi$ & $\tilde{\xi}$ & $\varphi_0^T$  & $\varphi_0^S$ & $\xi_0$\\
\hline
$A_4$ & $3$ & $1$ & $1$ & $1$ & $1$ &$1$ &$3$ & $1'$ & $3$ & $1$ & $1$ & $3$ &  $3$ & $1$\\
\hline
$Z_3$ & $\omega$ & $1$ & $1$ & $1$ & 1 & $\omega^2$&
$1$ & $1$ & $\omega$ & $\omega$ & $\omega$ & $1$ &  $\omega$ & $\omega$\\
\hline
$Z'_3$ &$1$ & $\omega$ & $\omega^2$ & $1$ & $1$ & $\omega^2$&
$\omega$ & $\omega$ & $1$ & $1$ & $1$ & $\omega$ &   $1$ & $1$\\
\hline
$U(1)_R$ & $1$ & $1$ & $1$ & $1$ & $0$ & $0$ & $0$& $0$  & $0$ & $0$ & $0$ & $2$ & $2$ & $2$\\
\hline
\end{tabular}
\caption{The transformation properties of leptons, electroweak Higgs doublets and flavons under $A_4 \times
Z_3 \times Z'_3$ and $U(1)_R$~.}
\label{transform}
\end{table}

\section{Vacuum Alignment}
In this section we will discuss the minimization of the scalar potential.
To achieve the desired alignment in a simple way, we work with a supersymmetric model,
with N=1 SUSY, eventually broken by small soft breaking terms.
In association with $\varphi_T$
we introduce another scalar field $\xi' \sim 1'$ under $A_4$.
In general, it is not easy to realize non-trivial minima for flavon fields
preserving different subgroups of $A_4$.
The key point is that the abelian part of the discrete symmetry $G$
should forbid unwanted terms in the driving superpotential.
In our case, the charged lepton hierarchy
is generated by $\langle \varphi_T \rangle \propto (0, 1, 0)$ 
which breaks entirely $A_4$. In the neutrino sector, the TB mixing pattern requires 
$\langle \varphi_S \rangle \propto (1, 1, 1)$ and $\langle \xi \rangle \ne 0$
which preserve a $Z_2$ subgroup of $A_4$.
We find that the minimal choice of $G$ in order to obtain the required 
vacuum alignment is given by $Z_3 \times Z'_3$. 

Apart from the discrete symmetry group $A_4 \times G$
the superpotential $w$ is automatically invariant also under a continuous $U(1)_R$
symmetry under which matter fields have $R=+1$, while Higgses and flavons have $R=0$. 
Such a symmetry will be eventually broken down to R-parity
by small SUSY breaking effects that can be neglected in the first approximation in our analysis. 
The spontaneous breaking of $A_4$ can be employed by introducing a new set of multiplets,
the driving fields, with $R=2$.
We introduce a driving field $\xi_0$, fully invariant under $A_4$, and two driving fields
$\varphi_0^T$ and $\varphi_0^S$, triplets of $A_4$. As
reported in Table \ref{transform},
$\xi_0$ and  $\varphi_0^S$ are charged under $Z_3$ which are responsible 
for the alignment of $\varphi_S$,
while $\varphi_0^T$ is charged under $Z'_3$ which drives a non-trivial VEV of $\varphi_T$.
The most general driving superpotential $w_d$ invariant under $A_4 \times G$ with $R=2$ is given by
\bea
w_d&=&g_1 (\varphi_0^S \varphi_S\varphi_S)+
g_2 \tilde{\xi} (\varphi_0^S \varphi_S)+
g_3 \xi_0 (\varphi_S\varphi_S)+
g_4 \xi_0 \xi^2+
g_5 \xi_0 \xi \tilde{\xi}+
g_6 \xi_0 \tilde{\xi}^2 \label{wd1}\\
&+& 
h_1 \xi' (\varphi_0^T \varphi_T)''+
h_2 (\varphi_0^T \varphi_T\varphi_T)~.\label{wd2}
\eea
Terms with $\varphi_S$, $\varphi_T$ interchanged 
are forbidden by $Z_3 \times Z'_3$. It is important to observe that the absence of 
$M (\varphi_0^T \varphi_T)$ is essential for the desired vacuum alignment of $\varphi_T$.
Eq.~(\ref{wd1}) and Eq.~(\ref{wd2}) give two decoupled sets of F-terms 
for driving fields which characterize the supersymmetric minimum.
From Eq.~(\ref{wd1}) we have:
\bea
\frac{\partial w}{\partial \varphi^S_{01}}&=&g_2\tilde{\xi} {\varphi_S}_1+
2g_1({\varphi_S}_1^2-{\varphi_S}_2{\varphi_S}_3)=0\nn\\
\frac{\partial w}{\partial \varphi^S_{02}}&=&g_2\tilde{\xi} {\varphi_S}_3+
2g_1({\varphi_S}_2^2-{\varphi_S}_1{\varphi_S}_3)=0\nn\\
\frac{\partial w}{\partial \varphi^S_{03}}&=&g_2\tilde{\xi} {\varphi_S}_2+
2g_1({\varphi_S}_3^2-{\varphi_S}_1{\varphi_S}_2)=0\nn\\
\frac{\partial w}{\partial \xi_0}&=&
g_4 \xi^2+g_5 \xi \tilde{\xi}+g_6\tilde{\xi}^2
+g_3({\varphi_S}_1^2+2{\varphi_S}_2{\varphi_S}_3)=0
\eea
We can enforce $\langle\tilde{\xi}\rangle=0$ 
{\footnote {Since there is no fundamental distinction between the singlets
$\xi$ and $\tilde{\xi}$ we have defined $\tilde{\xi}$ as the combination
that couples to $(\varphi_0^S \varphi_S)$ in the superpotential $w_d$.
The introduction of an additional singlet is essential to recover a non-trivial solution.}}
by adding to the scalar potential a soft SUSY breaking mass term
for the scalar field $\tilde{\xi}$, with $m^2_{\tilde{\xi}}>0$. In this case, in a finite portion of the parameter space, we find the solution
\begin{eqnarray}
\langle \tilde{\xi} \rangle&=&0~,~~~~~~~~~~~~\langle \xi \rangle=u~,\nn\\
\langle \varphi_S \rangle &=&(v_S,v_S,v_S)~,~~~~~~~~~v_S^2=-\frac{g_4}{3 g_3} u^2~,
\label{solS}
\end{eqnarray}
with $u$ undetermined, which is exactly analogue to the AF model. 
Setting to zero the F-terms from Eq.~(\ref{wd2}), we have:
\bea
\frac{\partial w}{\partial \varphi^T_{01}}&=&h_1\xi' {\varphi_T}_3+
2h_2({\varphi_T}_1^2-{\varphi_T}_2{\varphi_T}_3)=0\nn\\
\frac{\partial w}{\partial \varphi^T_{02}}&=&h_1\xi' {\varphi_T}_2+
2h_2({\varphi_T}_2^2-{\varphi_T}_1{\varphi_T}_3)=0\nn\\
\frac{\partial w}{\partial \varphi^T_{03}}&=&h_1\xi' {\varphi_T}_1+
2h_2({\varphi_T}_3^2-{\varphi_T}_1{\varphi_T}_2)=0\nn
\eea
and the solution to these four equations is:
\be
\langle \xi' \rangle =u' \ne 0~,~~~~~~~\langle \varphi_T \rangle =(0, v_T,0)~,~~~~~~~v_T=-\frac{h_1u'}{2h_2}~.
\label{solT}
\ee
with $u'$ undetermined. The flat directions can be removed by the interplay of radiative corrections to the scalar potential and soft SUSY breaking terms for $\xi'$, with $m^2_{\xi'}<0$.
It is worth to observe that since the two abelian factors in $Z'_3 \times Z_3$ 
are complementary for ($\varphi_T, \xi'$) and ($\varphi_S, \xi$), the VEV alignments
(\ref{solS}) and (\ref{solT}) are independent up to $1/\Lambda^2$.

\section{A $A_4 \times Z_3 \times Z'_3$ model for leptons}

In this section we propose a very simple SUSY model for leptons 
based on the following pattern of symmetry breaking of $A_4 \times Z'_3 \times Z_3$
\be
\begin{array}{c}
\langle\varphi_T\rangle=(0,v_T,0)~~~,~~~~~\langle\varphi_S\rangle=
(v_S,v_S,v_S)~~~\\[5pt]
\langle\xi\rangle=u~~~,~~~~~\langle\tilde{\xi}\rangle=0~~~,~~~~~\langle\xi'\rangle=u'~~~.
\end{array}
\label{vevs}
\ee
In the charged lepton sector the flavour symmetry $A_4 \times Z'_3$ is broken by 
$(\varphi_T,\xi')$ down to $G_{T^+}$ (with $Z_3$ unbroken) at leading order where
only the tau mass is generated.
The muon and electro masses are generated by higher order contributions.
In the neutrino sector $A_4 \times Z_3 $ is broken by $(\varphi_S,\xi)$ down to $G_S$ 
(with $Z'_3$ unbroken) with an 
accidental extra $G_{2-3}$ symmetry as noted benfore.

Precisely, the lepton masses are described by $w_l$, given by, up to $1/ \Lambda^3$:
\bea
w_l &=& \alpha_1 \tau^c (l \varphi_T)h_d/\Lambda \nn \\
&+& \beta_1 \mu^c \xi' (l \varphi_T)'' h_d/\Lambda^2+\beta_2 \mu^c (l \varphi_T \varphi_T) h_d/\Lambda^2 \nn \\
&+& \gamma_1 e^c (\xi')^2 (l \varphi_T)' h_d/\Lambda^3+
\gamma_2 e^c \xi' (l \varphi_T \varphi_T)'' h_d/\Lambda^3+
\gamma_3 e^c (l \varphi_T \varphi_T \varphi_T)h_d/\Lambda^3 \nn \\
&+& \gamma' e^c (l \varphi_S ) \xi^2 h_d/\Lambda^3 +\gamma'' e^c (l \varphi_S \varphi_S \varphi_S) h_d/\Lambda^3 \nn \\
&+&(x_a\xi+\tilde{x}_a\tilde{\xi}) (ll)h_u h_u/\Lambda^2+x_b (\varphi_S ll)h_u h_u/\Lambda^2+ \dots
\label{wlplus}
\eea
where the forth line is due to the interaction between charged lepton and neutrino sectors.
After electroweak symmetry breaking, $\langle h_{u,d}\rangle=v_{u,d}$~, 
given the specific orientation of 
$\langle \varphi_T \rangle \propto (0, 1, 0)$~, the first three lines in
$w_l$ give rise to diagonal and hierarchical mass terms for charged leptons.
The only off diagonal contributions to the charged lepton mass matrix come
from the fourth line in Eq.~(\ref{wlplus}) after electroweak and flavour symmetry breakings.
The off diagonal entries can be however rotated away by small unobservable right-handed
rotations (almost of order $O($VEV$/\Lambda)$)
without affection eigenvalues of $m_l$ up to $1/ \Lambda^3$ and
we will not consider these terms in the following.
Defining the expansion parameter $v_T/\Lambda \equiv \lambda^2 \ll 1$
and the Yukawa couplings $y_l$ ($l=e, \mu, \tau$) as 
\bea
y_\tau&=&\alpha_1 \nn~, \\ 
y_\mu&=& (\beta_1u'/v_T +2 \beta_2 ) \lambda^2\nn ~, \\
y_e &=& (\gamma_1(u'/v_T)^2-\gamma_2 u'/v_T-2\gamma_3+\gamma' v_S u^2/v^3_T) \lambda^4 \nn~, 
\eea
the charged lepton masses are given by
\be
m_l=y_l \lambda^2 v_d~~~~~~~(l=e,\mu,\tau)~~~.
\ee
Here we assume that all VEVs of flavon fields are of the same order of magnitude.
Differently from the original proposal of $A_4$ \cite{TB1, TB2} model or $T'$ model of \cite{TB3}, 
we are able to produce the required hierarchy among 
$m_e$, $m_\mu$ and $m_\tau$ provided $\lambda \approx \lambda_c$ 
where $\lambda_c$ is the Cabibbo angle. 
Some observations are in order. The different assignment of lepton singlets 
$e^c$, $\mu^c$ and $\tau^c$ under $Z'_3$ needs to compensate the $Z'_3$
charge of $\varphi_T$.
As a result, we obtain automatically a diagonal $m_l$.
However the alignment (\ref{vevphiT}) is, in our model, 
the true fundamental ingredient in generating the charged lepton hierarchy.
As already mentioned, the two abelian factors $Z'_3 \times Z_3$ 
are complementary for the charged lepton
and neutrino sectors. This means that the interaction between the two sectors
can arise only at a relative order $1/ \Lambda^3$. 
The only field which is charged under both abelian factors is $h_d$.
As we will explain later on, the previous almost diagonal structure
of charged leptons might receive important off-diagonal contributions  only 
when additional breaking effects of $A_4$ are added.

At this order, all the information about lepton mixing angles is encoded in the neutrino mass matrix 
which is identical to one of the original $A_4$ model and $T'$ model:
\be
m_\nu=\frac{v_u^2}{\Lambda}\left(
\begin{array}{ccc}
a+2 b& -b& -b\\
-b& 2b& a-b\\
-b& a-b& 2 b
\end{array}
\right)~~~,
\label{mnu0}
\ee
where
\be
a\equiv x_a\frac{u}{\Lambda}~~~,~~~~~~~b\equiv x_b\frac{v_S}{\Lambda}~~~.
\label{ad}
\ee
The neutrino mass matrix is diagonalized by the transformation:
\be
U^T m_\nu U =\frac{v_u^2}{\Lambda}{\tt diag}(a+3b,a,-a+3b)~~~,
\ee
with $U=U_{\text{TB}}$~.
Therefore the TB mixing of eq. (\ref{TB}) is reproduced, at the leading order.
For the neutrino masses we obtain:
\bea
|m_1|^2&=&\left[-r+\frac{1}{8\cos^2\Delta(1-2r)}\right]
\Delta m^2_{atm}\nn\\
|m_2|^2&=&\frac{1}{8\cos^2\Delta(1-2r)}
\Delta m^2_{atm}\nn\\
|m_3|^2&=&\left[1-r+\frac{1}{8\cos^2\Delta(1-2r)}\right]\Delta m^2_{atm}~~~,
\label{lospe}
\eea
where $r\equiv \Delta m^2_{sol}/\Delta m^2_{atm}
\equiv (|m_2|^2-|m_1|^2)/(|m_3|^2-|m_1|^2)$,
$\Delta m^2_{atm}\equiv|m_3|^2-|m_1|^2$ 
and $\Delta$ is the phase difference between
the complex numbers $a$ and $b$. 
The value of $|m_{ee}|$, the parameter characterizing the 
violation of total lepton number in neutrinoless double beta decay,
is given by:
\be
|m_{ee}|^2=\left[-\frac{1+4 r}{9}+\frac{1}{8\cos^2\Delta(1-2r)}\right]
\Delta m^2_{atm}~~~.
\ee
Independently from the value of the unknown phase $\Delta$
we get the relation:
\be
|m_3|^2=|m_{ee}|^2+\frac{10}{9}\Delta m^2_{atm}\left(1-\frac{r}{2}\right)~~~,
\label{pred1}
\ee
which is another prediction of our model. 
The results on light neutrinos coincide with those obtained at leading order in the $A_4$ and $T'$ 
models \cite{TB2, TB3}.
The new feature of our model is that the enlarged abelian symmetry $G$ 
strongly suppresses possible higher order contributions and
the prediction of the TB mixing pattern and Eq.~(\ref{pred1}) is (almost, i.e. up to terms with relative suppression of order $1/\Lambda^2$) exact.

\section{Deviations from TB mixing}

The results of the previous sections hold almost exactly.
As already pointed out, the $Z_3 \times Z'_3$ charge assignments in Table \ref{transform} forbid all
next-to-leading and next-to-next-to-leading order corrections.
The non-zero higher order corrections arise only at the relative order $1/\Lambda^3$
with respect to the terms already considered in $w_d$ and $w_l$.
In particular, our model predicts the almost exact TB mixing in the neutrino sector.
From the phenomenological point of view, it is also interesting to 
explore a natural mechanism to generate sizable 
deviations from TB mixing without loosing the predictivity.

In the literature, model independent approaches to parametrize small corrections
to the TB mixing pattern have been widely developed.
We should not re-analyze fully all these issues in our model.
Particular emphasis concerns special patterns of TB breaking effects, 
both for the neutrino and the charged lepton sectors,
that can lead to testable correlations between parameters \cite{dev}.
Here we will focus on the so called neutrino mixing sum rule given
by Eq.~(\ref{sumrule}) 
mentioned in the introduction. The goal is to provide an existence proof that predictive
deviations from TB mixing can be generated naturally in our constrained $A_4$ model.

\subsection{$\theta^e_{13}$ ($\theta^e_{12}$) dominance}
One of the most interesting deviations from TB mixing 
is called $\theta^e_{13}$ ($\theta^e_{12}$) dominance  \cite{dev}.
This pattern is realized by assuming that, in the neutrino sector,
$U_\nu=U_\text{TB}$ holds exactly and the
only corrections come from a non trivial $U_e$.
In this case, the unitary matrix $U_e~$, parametrized in the standard way, 
shall involve three small rotations $\theta_{12}^e$, $\theta_{13}^e$, $\theta_{23}^e~$ 
and in general three phases.
From $U_{PMNS}=U_e^\dagger U_\nu$, one can show
that, at first order, the resulting $\theta_{13}$ and $\theta_{12}$ 
depend only on $\theta^e_{13}$, $\theta_{12}^e$ and $\delta$, 
a combination of the phases which appear in $U_e$
(for more detail, see \cite{dev}). 
If the $\theta^e_{13}$ ($\theta^e_{12}$) contribution 
dominates over the $\theta^e_{12}$ ($\theta^e_{13}$) one, in the literature referred also as
$\theta^e_{13}$ ($\theta^e_{12}$) dominance,
one obtains the following TB mixing sum rule:
\be
\text{sin}^2\theta_{12}=\frac13+\frac{2\sqrt{2}}{3}\text{cos} \delta ~\text{sin}\theta_{13}~~~.
\label{pred2}
\ee
To my present knowledge, a model realization of the previous prediction, based on symmetry principle,
has never been considered before.
 Indeed, (\ref{pred2}) is not a natural prediction of the original AF model in which
$\theta^e_{12} \sim \lambda^2$ and $\theta^e_{13} \sim \lambda^2$ are
generated at the same time, together with the
subleading contributions to the neutrino sector, all of order $\lambda^2$.
Differently, in our constrained $A_4$ model, an exact TB mixing in the neutrino sector
with a diagonal and hierarchical charged lepton mass matrix is predicted without corrections 
(up to $1/\Lambda^2$).
Only additional scalar fields should account for deviations from TB mixing.
We will propose a simple extension of our model in which, for example, the 
$\theta^e_{13}$ dominance pattern arises from spontaneous breaking of
$A_4$ with an additional $A_4$ singlet.

\subsection{$\theta^e_{13}$ dominance in the constrained $A_4$ model}

The leading order results of our model can be slightly modified 
by introducing another singlet $\chi$, invariant under $Z_3$ 
but carrying a charge $\omega^2$ under $Z'_3$.
As shown in the Appendix A, if $\chi$ acquires a large VEV $\langle\chi\rangle=v \sim u'$, 
it can lead to a sizable correction to $\langle \varphi_T \rangle = v_T(0,1,0)$.
At the first order in $1/\Lambda$, only the first component of $\langle \varphi_T \rangle$
receives the correction. This is a new feature of the constrained
$A_4$ model. As we shall see in a moment, this result is essential 
to reproduce the $\theta^e_{13}$ dominance pattern
{\footnote{If $\chi$ were $\chi \sim (1',1,\omega^2)$ 
under $A_4 \times Z_3 \times Z_3'$, we would cover the $\theta^e_{12}$ dominance pattern.}}.

Now we begin to discuss the effect of $\chi$ on lepton masses and mixing.
The introduction of $\chi$ slightly modifies $w_l$ of Eq.~(\ref{wlplus}):
\be
w_l \rightarrow w_l+\beta_3 e^c \chi (l \varphi_T) h_d/\Lambda^2+\gamma_4 \mu^c \chi^2 (l \varphi_T) h_d/\Lambda^3~~~.
\ee
Taking into account these new terms in the superpotential and 
the subleading correction to the VEV of $\varphi_T$
the charged lepton mass matrix becomes (up to $1/ \Lambda^3$):
\be
m_l=\left(
\begin{array}{ccc}
O(y_e) &  0 & \beta_3 \lambda^2  \\
0 & y_\mu & O(y_e) \\
\alpha_1 \delta v_T/(\Lambda \lambda^2) & 0  & y_\tau
\end{array}
\right) \lambda^2 v_d~~~,
\label{ml1}
\ee
where we explicitly give only terms of order up to $\lambda^4$.
In our notation, the transformation needed to diagonalize $m_l$ is:
\be
V_e^T m_l U_e=diag(m_e,m_\mu,m_\tau)~~~
\ee
and the unitary matrix $U_e$ involves rotations of order
$\theta_{12}^e=0$, $\theta_{13}^e=O(\lambda^2)$, $\theta_{23}^e=0$. 
Recalling that the neutrino mass matrix $m_\nu$ does not
receive corrections up to those of relative order $O(\lambda^6)$, 
$U_{\nu}$ is given by Eq.~(\ref{TB}) almost exactly. 
Then we realize naturally the $\theta^e_{13}$ dominance pattern
without adjusting adimensional coefficients. 
Indeed, in this simple extension, we not only reproduce the sum rule (\ref{pred2})
but also
$$ \text{sin}^2\theta_{13}=O(\lambda^2)~.$$

The subleading effect considered in this section is very important 
in order to incorporate CP violation in the TB pattern. 
The smallness of $\theta_{13}$ and CP violation are correlated in the
$\theta^e_{13}$ dominance pattern \cite{dev}.
The size of CP violation in the neutrino oscillations is measured by the following
Jarlskog invariant \cite{Jarlskog}:
$$J_\text{CP}= \text{Im}\{U_{e1}U_{\mu 2}U^*_{e2} U^*_{\mu 1}\}~,$$
depending on the Dirac phase $\delta$.
In our case, we obtain a very simple expression:
$J_{CP} \approx \frac16  \lambda^2 ~\text{sin} \delta$.
Since $\text{sin}^2\theta_{12}$ is very close to $1/3$, a value of $\theta_{13}$ 
close to the existing upper limit, necessarily requires a large CP violation.
However, without accidental enhancement, we typically have a small $\theta_{13}$, say at a
level of $O(\lambda_c^2)$. In this case, the value of $\delta$ is very sensible to
the closeness of $\text{sin}^2\theta_{12}$ to its tri-bimaximal value.
If a sizable CP violation is determined by long-baseline neutrino oscillation experiments,
we predict a $\text{sin}^2\theta_{12}$ to be extremely close to $1/3$.
Conversely, if a future high precision determination of
$\text{sin}^2\theta_{12}$ will come extremely close to $1/3$, 
we should expect an observable CP asymmetry in long-baseline neutrino oscillation experiments.

\section{Further discussions and conclusion}

Both charged fermion mass hierarchies and large lepton mixings can be potentially 
achieved via spontaneous breaking of the flavour symmetry.
However, in most cases, the flavour group is of the type $D\times U(1)_{FN}$ where
$D$ is a discrete component that controls the mixing angles and $U(1)_{FN}$ is an 
abelian continuous symmetry that describes the mass hierarchies.
It would be a very attractive task to construct economical and constrained models
where the same flavon fields producing the mixing pattern by VEV misalignment 
are also responsible for the mass hierarchies.
Models of this type based on the gauged flavour groups $SU(3)$,
$SO(3)$ \cite{su3} and the non-abelian finite group $PSL_2(7)$ \cite{largediscrete} exist in the literature. 
However in these models the charged fermion mass hierarchies are obtained
by a complicated flavour symmetry breaking sector and/or an ad hoc adjustment of the messenger
scales. A first simple model belonging to this class has been constructed in \cite{S3}
based on the minimal non-abelian group $S_3$.
With a very economical flavon sector, the model is able to provide a decent description 
of the main features of all the fermion mass spectrum, including the approximate vanishing of $\theta_{13}$ and of $\theta_{23}-\pi/4$.

In the present work, we addressed a possible unified picture of the charged lepton mass hierarchy 
and the Tri-Bimaximal mixing pattern. 
The model is based on the spontaneously broken $A_4$ flavour symmetry
with an abelian factor given by $Z_3 \times Z'_3~$.
The very special structure of the leptonic mixings is 
understood by a mechanism of vacuum misalignment in flavour space.
A new type of SUSY minima of the scalar potential is explicitly given.
In the neutrino sector, the $A_4$ component is broken to $G_S \times G_{2-3}$
(where $G_{2-3}$ is an accidental symmetry) as in the original AF models
guaranteeing the TB mxing.
On the other hand, in the charged lepton sector, $A_4$ is entirely
broken already at the leading order of the $\langle \varphi \rangle/ \Lambda$ expansion.
A noticeable feature of the model is that the charged lepton hierarchy is also
determined by the symmetry breaking of the $A_4$ group
without the use of an extra $U(1)_{FN}$ component.

The lepton mixing data have been undergone a surprising improvement in the last years. 
From the phenomenological point of view, suitable flavour models should 
be subject of precise tests.
It would be desirable to have some criteria to discard existing models by experiments.
For this reason, another goal of the present work is to construct a constrained $A_4$ model
for TB mixing leading to testable sum rules. Many flavour models
depend on the smallness of $\theta_{13}$. However, nowadays a sensitivity of
a $\lambda_c^2$ level is reached only for $\theta_{12}$.
Constraining the angles  $\theta_{13}$ and $\theta_{23}$ by a similar sensitivity
will require some more years of work.
There are two different classes of neutrino flavour models depending on 
if $\theta_{13}$ is not so small, say of $O(\lambda_c)$, or $\theta_{13}$ is very small, say of $O(\lambda^2_c)$. Our model belongs to the second class of models. 
Furthermore, the predicted
value of $\theta_{13}$ is subjected to the TB sum rule:
$\text{sin}^2\theta_{12}=1/3+2\sqrt{2}~(\text{cos} \delta ~\text{sin}\theta_{13})/3~.$
We find that this is a general feature of our model:
possible deviations from TB mixing are highly constrained giving rise to
very stringent next-to-leading predictions.

In the end, we are left with the question whether our model could provide a satisfactory
description of the quark sector as well.
The most naive way to include quarks is to adopt for them the same classification scheme 
under $A_4\times G$ that we have used for leptons.
Proceeding in this way, we would obtain an approximately diagonal $V_{CKM}$.
This is a good first order result for quark sector. 
However, since the mass matrices of up- and down-type quarks
belong to the same assignment, one should wonder if 
they were diagonalized by the same unitary matrix leading to a diagonal $V_\text{CKM}$,
even including subleading corrections.
This is the case, indeed, in a large class of $A_4$ models \cite{A4-quarks}
and the common opinion is to introduce additional $A_4$ breaking effects.
On the other hand, undesirable sources of $A_4$ breaking would be dangerous since
they could destabilize the desired vacuum alignment.
Differently, in our construction, additional sources of $A_4$ breaking
can be easily kept under control and they could play a role in generating
a non-trivial $V_{CKM}$.
Nonetheless there is another lack if we were adapt the the same field assignment 
under $A_4 \times G$ that we have used for leptons also for quarks.
The top mass should be generated just at a non-renormalizabel level 
and this is not a good feature.
A detailed analysis of these issues is beyond the scope of this paper and we will 
leave if for a future work.

\vskip 0.5 cm
\section*{Acknowledgements}
We thank Ferruccio Feruglio for useful suggestions and for his encouragement in our work.
We thank also Claudia Hagedorn and Luca Merlo for useful discussions 
and for reading the preliminary manuscript of the work.
We recognize that this work has been partly supported by 
the European Commission under contracts MRTN-CT-2004-503369 and MRTN-CT-2006-035505.

\newpage
\section*{Appendix A}

In this appendix we will study the effect of $\chi \sim (1,1,\omega^2)$ 
under $A_4 \times Z_3 \times Z_3'$ to the vacuum alignment.
All the leading order operators displayed in $w_d$, see Eq.~(\ref{wd1}-\ref{wd2}), 
are of dimension three. 
With the introduction of $\chi~$, the driving superpotential should be modified into $$w_d \rightarrow w_d +\Delta w_d~.$$
Given the charge assignment of $\chi$, $\Delta w_d$ contains only one
additional operator up to dimension four:
\be
t_1 (\varphi^T_0 \varphi_T) \chi^2~.
\ee
After inclusion of this operator, the minimum 
for $\varphi_T$, $\xi'$, $\varphi_S$, $\xi$, $\tilde{\xi}$ should be shifted.
One can look for the new minimum by a perturbation expansion
in the power of $1/\Lambda$ around the leading order VEVs given in Eq.~(\ref{vevs}).

We find that, at the first order in $1/\Lambda$, all the leading order VEVs
remain unchanged except for $\varphi_T$:
\be
\begin{array}{c}
\langle\varphi_T\rangle=(\delta v_{T1},v_T,0)~~~,~~~~~\langle\varphi_S\rangle=
(v_S,v_S,v_S)~~~\\[5pt]
\langle\xi\rangle=u~~~,~~~~~\langle\tilde{\xi}\rangle=0~~~,~~~~~\langle\xi'\rangle=u'~~~,~~~~~\langle\chi\rangle=v~~~,
\end{array}
\label{hovevs}
\ee
where $\delta v_{T1}= xv^2/\Lambda$ with $x= t_1/4h_2$ and $u, u', v$ are undetermined.
$v$ can slide to a large scale if we introduce a soft mass term with $m^2_{\chi} < 0$.
However, in order to make sense the previous perturbative analysis, we have to require
$v/\Lambda \lesssim \lambda^2$.
As expected, a sizable shift $\delta \varphi_{T1}$ can be generated by a relatively large
VEV of $\chi$. We will require $v/\Lambda\approx \lambda^2$ 
and this fixes the order of magnitude of the 
subleading corrections:
\be
\frac{\delta v_{T1}}{\Lambda} \approx \lambda^4~~~.
\label{dvev}
\ee
From this analysis we learn that the order of suppression of 
$\delta \text{VEV}$ depends crucially on the abelian factor $G$
and its expected values should be (\ref{dvev}) but do not necessarily lie in this range.
Different components could in principle receive corrections with different suppressions,
even though all leading flavon VEVs are assumed to be of the same order.
A more detailed perturbation analysis, in fact, shows that $\delta v_{T3} \sim \lambda^2 
\delta v_{T1}$.

To summarize, the introduction of extra singlets perturbs the leading order alignment of 
$\varphi_T$.
Due to the presence of the second $Z'_3$ factor, an internal hierarchy appears in the VEV structure 
of the triplet $$\langle\varphi_T\rangle=v_T(\lambda^2,1,\lambda^4).$$
This is another nice feature of our constrained $A_4$ model.
The previous perturbation analysis shows a simple example in which 
a non-abelian symmetry can be fully compatible with FN mechanism with abelian charges.

\end{document}